# Convolutional Neural Networks for Reflective Event Detection and Characterization in Fiber Optical Links Given Noisy OTDR Signals

Khouloud Abdelli
Advanced Technology & Chair of Communications
ADVA & Kiel University
82152 Munich, Germany
kabdelli@adva.com

Helmut Grießer
Advanced Technology
ADVA Optical Networking SE
82152 Munich, Germany
HGriesser@adva.com

Stephan Pachnicke
Chair of Communications
Kiel University
24143 Kiel, Germany
stephan.pachnicke@tf.uni-kiel.de

*Abstract*— Fast and accurate fault detection and localization in fiber optic cables is extremely important to ensure the optical network survivability and reliability. Hence there exists a crucial need to develop an automatic and reliable algorithm for real-time optical fiber faults' detection and diagnosis leveraging the telemetry data obtained by an optical time domain reflectometry (OTDR) instrument. In this paper, we propose a novel data-driven approach based on convolutional neural networks (CNNs) to detect and characterize the fiber reflective faults given noisy simulated OTDR data, whose SNR (signal-to-noise ratio) values vary from 0 dB to 30 dB, incorporating reflective event patterns. In our simulations, we achieved a higher detection capability with low false alarm rate and greater localization accuracy even for low SNR values compared to conventionally employed techniques.

*Keywords—fiber fault diagnosis, optical time domain reflectometry, machine learning, long short-term memory*

## I. INTRODUCTION

Real-time fiber link monitoring and diagnosis is of crucial importance. It enables to quickly discover and pinpoint the faults in fiber optics and thereby helps to reduce operation-and-maintenance expenses (OPEX), to minimize the mean time to repair (MTTR) and to enhance the network quality. Fiber optic link monitoring has been widely performed using optical time domain reflectometry (OTDR), an optoelectronic instrument commonly used to measure fiber characteristics and to detect and locate cable faults. The OTDR operates like an optical radar. By injecting a series of optical pulses into the fiber under test, these pulses are partially reflected and scattered back towards the source due to the Rayleigh scattering. The strength of the reflected signals is recorded as a function of propagation time of the light pulse, which can be converted into the length of the optical fiber [1]. Consequently, with the recorded OTDR trace (or waveform) the positions of the faults, including fiber misalignment/mismatch, fiber breaks, angular faults, dirt on connectors and macro-bends [2] along the fiber can be identified. These events can be broadly categorized as either reflective or non-reflective. The reflective events are Fresnel peaks and result from sudden changes in the density of the material, usually fiber-to-air transitions. The non-reflective events are results of mode field diameter (MFD) variations caused by geometric changes or differences in the glass fiber and lead to attenuation but no reflection [3].

Analyzing OTDR traces can be tricky, even for experienced field engineers, mainly due to the noise overwhelming the signal, leading to inaccurate or unreliable event detection and localization. Averaging multiple OTDR measurements helps to reduce the noise and thus to improve the performance of OTDR event analysis approaches in terms of event detection and fault localization accuracy. However, the averaging process is time consuming. As the optical fiber real-time detection is considered as the industry standard, it is essential to have an automated reliable technique that detects and locates events in a timely manner and with high accuracy while processing noisy OTDR data without the need to perform long averaging and without requiring support by trained personnel.

Conventionally, the OTDR event analysis technique is based on a two-point method combined with the least square approximation technique that calculates the best fit line between two markers placed on the section of the OTDR trace to calculate the distance to the event and loss at an event (a connector or splice) between the two markers [4]. Although this method is very simple, it is coarse and noise-sensitive [5]. Some OTDR event detection and localization approaches based on either the Gabor transform [6] or the Wavelet analysis [1] have been proposed. However, they are either numerically complex or locate the faults inaccurately, particularly for OTDR signals with low SNR levels. Recently, data-driven approaches based on machine learning (ML) have shown great potential in processing fault diagnostics given sequential data. We have presented ML models for laser failure detection and prediction given noisy current sensor data [7]-[9].

In this paper, we propose a novel diagnostic model based on convolutional neural networks (CNNs) for fiber reflective event detection, localization, and characterization in terms of reflectance. The overview of the proposed approach is shown in Fig. 1. Our approach is applied to simulated OTDR data modelling the reflective fault patterns for different SNR values ranging from 0 to 30 dB. It takes as inputs the preprocessed OTDR sequences and predicts the characteristics of the identified event. The results show that the presented approach detects and locates the reflective events with higher accuracy compared to the conventional OTDR techniques thanks to the capability of the CNNs to learn and extract the optimal features underlying the reflective event pattern even at low SNR values.

The remainder of the paper is organized as follows. Section 2 describes the simulation setup for data generation and the architecture of the proposed approach. The results showing the performance evaluation of the proposed model as well as its comparison with a conventional OTDR approach are presented in Section 3. Conclusions are drawn in Section 4.

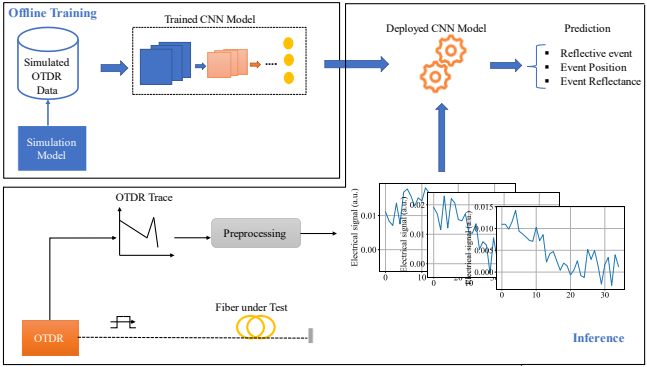

Fig. 1: Overview of the proposed approach with offline training and inference

## II. SETUP & CONFIGURATIONS

### A. Data Generation

To train the ML approach, synthetic OTDR data is generated. We simulate the operational principle of OTDR by modelling the different OTDR components like pulse generator, coupler, photodiode. A reflector is used to induce a reflective event in the fiber link. For our simulation, the SNR is considered as the relevant figure-of-merit since the different simulation parameters namely the laser power, the length of the fiber, the attenuation and the receiver photodiode sensitivity influence the SNR.

The OTDR simulation is conducted using VPIphotonics Transmission Maker. The data generation model is depicted in Fig. 2. The pulse generator module creates rectangular pulses of pulse width $T$ filtered by a Bessel low pass filter with a bandwidth of $\frac{0.6}{T}$. An EAM imprints the filtered pulses on the laser light. The laser and the modulator are assumed to be ideal. The optical pulses are then launched into the bidirectional fiber. By hitting the reflector (95% reflection), the signal is sent back to the photodiode that converts them to an electrical current. The model assumes that the photodiode is ideal (i.e. a simple squaring device) and that all noise is white Gaussian and added at the transimpedance amplifier (TIA) of the photodiode, resulting in a certain SNR value. The SNR (in dB) is defined as

$$\text{SNR} = 10 \log_{10} \frac{\text{reflective event energy}}{\text{noise energy at the event position}} \quad (1)$$

The fiber model is linear, assuming that the optical launch power is low. Due to the high reflection at the end of the fiber, Rayleigh scattering in the fiber is negligible and not considered in the fiber model. The pulse width $T$ is fixed to 100 ns, whereas the SNR is chosen from uniform distributions. 30,000 OTDR traces are generated. The OTDR signals are then segmented into fixed sliding windows of length 35. We randomly select from each segmented trace 8 sequences: 4 sequences containing no reflective event and 4 sequences including a part or the whole peak (i.e. reflective fault) pattern. In total, a data set comprised of 240,000 sequences is built. Our approach takes as input the sequence of signal power values and outputs the $Class_{id}$ (0: no reflective event, 1: reflective event), the reflective event position index within the sequence, and the reflectance $R$. The reflectance can be derived from the peak height compared to reference measurements for different cleaved fiber ends (see Section VI in [10]).

The generated data is normalized and divided into a 60% training dataset, a 20% validation dataset and a 20% test dataset.

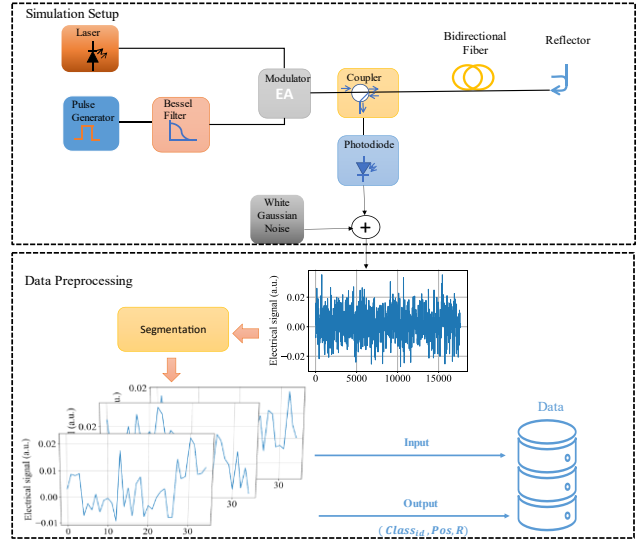

Fig. 2: Data generation process

### B. CNN based Model

#### a. Overview of CNN

CNNs are a type of artificial neural networks, biologically inspired by the modular structure of the human visual cortex. They have been widely used in computer vision and have become the state of the art for several object recognition tasks such as handwritten digit recognition. They are powerful at extracting features. The architecture of a CNN comprises different hidden layers namely the convolutional, the pooling and the fully connected layers, made by stacking them on top of each other in sequence. The convolutional layers are composed of filters, the neurons of the layer, and the feature maps. The output of a filter is applied to the previous layer and used to extract features. The pooling layers are used to down-sample the previous layers' feature map by consolidating the features learned and expressed in the previous layer's feature map. The fully connected layers are the normal feedforward neural network layer used to create final nonlinear combinations of features and for making predictions by the network.

#### b. Proposed CNN-based Model

The overall structure of the proposed model is depicted in Fig. 3. The architecture of our approach comprises mainly of 4 convolutional layers, a max pooling layer, a flattened layer and three fully connected feedforward layers defined for the three tasks namely reflective event detection $T_1$, event position estimation $T_2$ and reflectance prediction $T_3$. The convolutional layers contain 64, 32, 32 and 16 filters respectively followed by a dropout layer to avoid overfitting. The features extracted by the convolutional and max pooling layers are flattened into a one-dimensional vector before being transferred to the three fully connected feedforward layers composed of 16 neurons. The overall loss function used to update the weights of the model based on the error between the predicted and the desired output can be formulated as

$$Loss_{total} = \sum_{i=1}^{3} \lambda_i \cdot loss_{T_i} \qquad (2)$$

where $loss_{T_i}$ denotes the loss of task $T_i$, and the first task loss is the binary cross-entropy loss, whereas the other losses are regression losses (mean squared error). The loss weights $\lambda_i$ are hyperparameters to be tuned. For our experiments, the weight of each task loss is set to 0.33.

## III. RESULTS AND DISCUSSIONS

### A. ML Model Performance Evaluation

To evaluate the performance of the proposed model, several metrics were used including the detection rate (i.e. detection probability) ($P_d$) for the reflective event detection task evaluation and the root mean square error (RMSE) metric for the evaluation of the event position estimation and reflectance prediction tasks. $P_d$ is the portion of the total number of reflective events that were correctly detected. It is defined as follows:

$$P_d = \frac{TP}{TP + FN} \qquad (3)$$

where TP is the number of true reflective event detections, and FN is the number of false negative ones.

The false alarm probability ($P_{FA}$) is expressed as:

$$P_{FA} = \frac{FP}{FP + TN} \qquad (4)$$

where FP is the number of false positives, and TN is the number of true negatives.

Figure 4 shows the effects of the SNR on the reflective event detection capability of our approach for different levels of $P_{FA}$. As expected, $P_d$ increases with SNR. For SNR values higher than 13 dB, $P_d$ is approaching 1. For lower SNR values, the performance is worse as it is tricky to differentiate the reflective event from the noise and thereby the misclassification rate is higher.

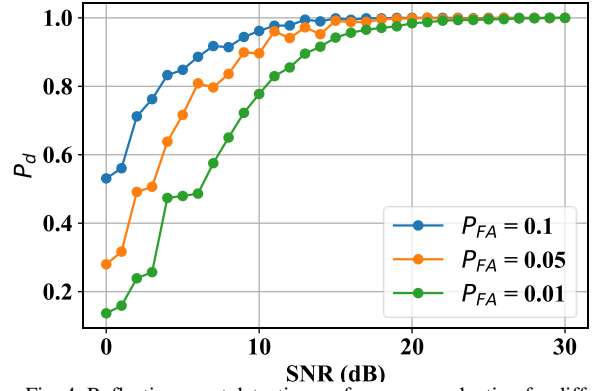

Fig. 4: Reflective event detection performance evaluation for different false alarm probability levels.

To investigate the influence of the SNR on the event localization accuracy, we evaluated the event position estimation capability of our model with three test datasets: one dataset containing the whole reflective event pattern, one dataset comprising just a part of the reflective event, and one dataset including the mix of the whole and partial reflective event patterns. The results of the evaluation are depicted in Fig. 5.

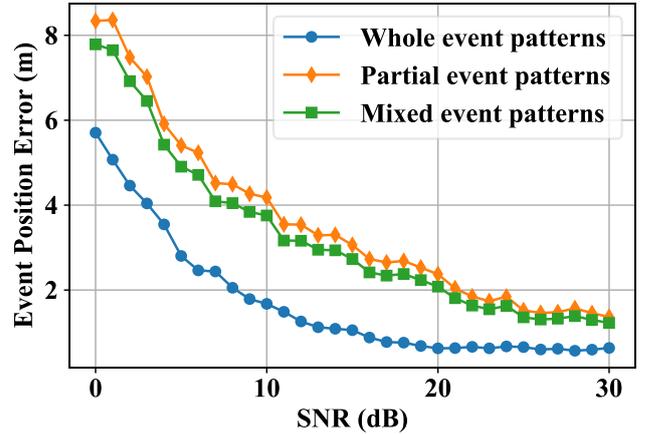

Fig. 5: Event position estimation error (RMSE) of the ML model

For the fed sequences incorporating the whole reflective event pattern, our approach locates the event with an RMSE less than 1 m for SNR values higher than 15 dB.

Whereas for the sequences containing partial event patterns,

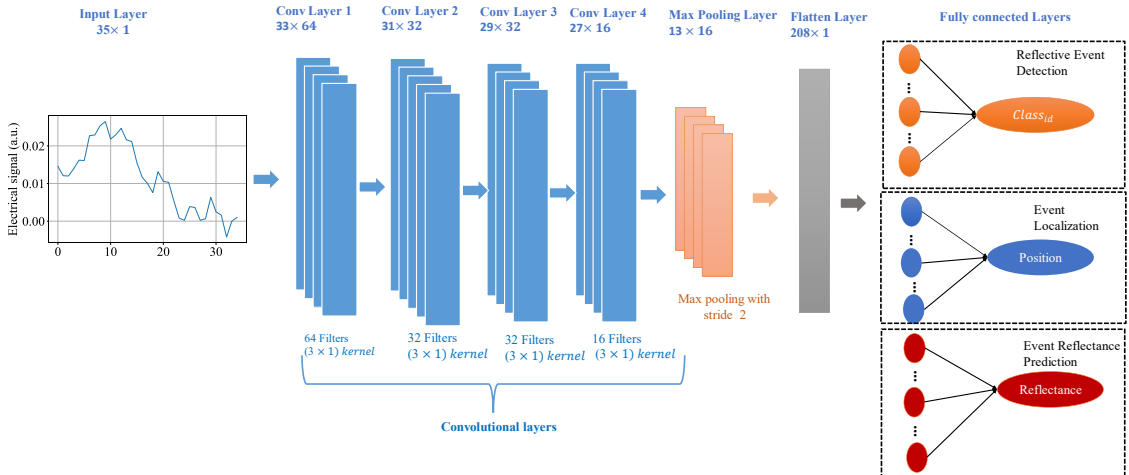

Fig. 3: Proposed approach structure

the RMSE is higher and it is less than 2 m for SNR values higher than 21 dB.

Figure 6 shows that the RMSE of the reflectance prediction decreases as SNR increases. For lower SNR values (SNR ≤10 dB), the RMSE is higher than 10 dB, and for higher SNR values, it can be as low as 3 dB.

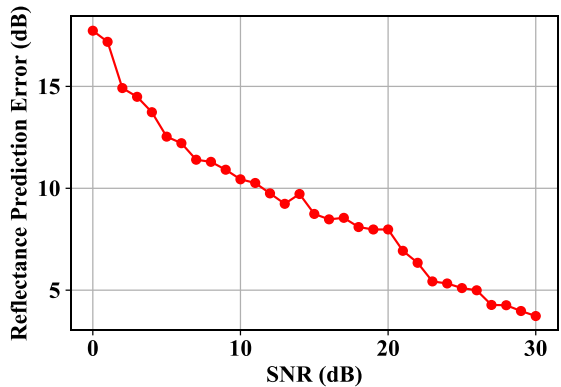

Fig. 6: Reflectance prediction error (RMSE) of the ML model

*B. ML Model versus Conventional method*

It should be noted that the developed ML model is trained with data including partial and whole peak sequences since the model should be applied to arbitrarily segmented OTDR sequences. For a fair comparison of the developed ML model with a conventional rank-1 matched subspace detector (R1MSDE [11]), an unseen test dataset, containing only the complete reflective event pattern or no event, was generated. R1MSDE uses the theory of matched subspace detection and associated maximum likelihood estimation procedures to distinguish connection splice events from noise and the Rayleigh component in the OTDR data. As in [11] the reflective event is modelled by a rectangular pulse with prior knowledge of the duration of the event. Given that the matched filter is optimum for unipolar modulation detection for a linear channel with additive Gaussian noise and that the optimum OTDR detector for single pulses (i.e. single reflective events) cannot be better due to the unknown peak position, the optimum unipolar modulation performance serves as an upper bound for an optimum peak detector. The detection rate $P_d$ for the case of optimum detection is calculated as follows:

$$P_d = \tfrac{1}{2}\text{erfc}\left(2(\delta-1)\sqrt{\tfrac{1}{2}SNR_{lin}}\right) \quad (5)$$

where erfc denotes the complementary error function, $SNR_{lin}$ is the linear SNR and $\delta$ is the detection threshold expressed as a function of $P_{FA}$

$$\delta = \frac{\text{erfcinv}(2 P_{FA})}{\sqrt{2 SNR_{lin}}} \quad (6)$$

A comparison of the different detectors in terms of $P_d$ for a $P_{FA}$ of 0.1, as shown in Fig. 7, demonstrates that the ML model outperforms the R1MSDE and is closer to the upper bound of the optimum detector.

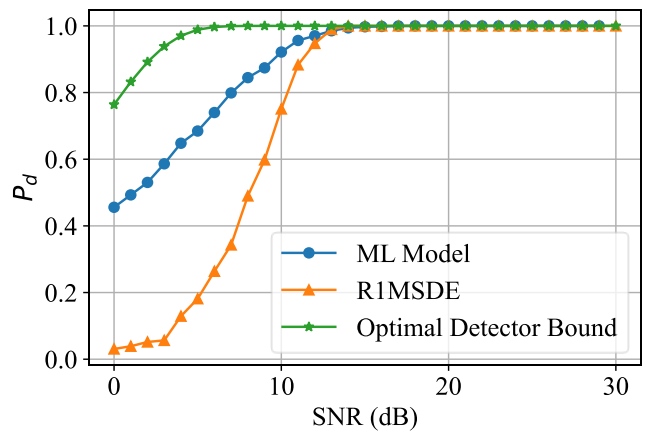

Fig. 7: Comparison of the detection rate for various algorithms.

The results of the comparison of the ML model performance with R1MSDE in terms of RMSE of the event position are shown in Fig. 8. For lower SNR values, the ML model outperforms the R1MSDE by achieving a smaller peak position error of less than 5 m. As SNR increases, the peak position error of R1MSDE is getting closer but is still worse than the proposed ML model.

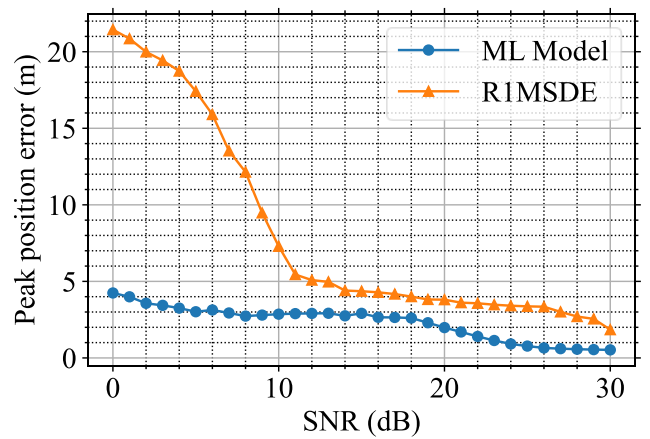

Fig. 8: Peak position estimation error (RMSE) for the ML model vs. R1MSDE

## IV. CONCLUSIONS

An ML model based on a CNN is proposed to detect and locate reflective events and estimate their reflectance given noisy OTDR data. The results showed that the ML model outperformed the conventional OTDR technique. Future work will include improving the reflectance prediction capability.